\documentclass[%
 reprint,
preprintnumbers,
amsmath,amssymb,
aps,
prl,
]{revtex4-2}

\usepackage{graphicx}
\usepackage{dcolumn}
\usepackage{bm}
\usepackage{mathtools} 
\usepackage{lipsum}
\usepackage{tikz-cd}
\usepackage{tikz}
\usepackage[caption=false]{subfig}
\usepackage{hyperref}

\newcommand{\id}{\textnormal{id}}

\newcommand{\Z}{\mathbb{Z}}

\newcommand{\R}{\mathbb{R}}
\newcommand{\C}{\mathbb{C}}

\newcommand{\I}{\mathrm{i}}

\begin{document}

\preprint{ZMP-HH/26-29}

\title{Elliptic spin Ruijsenaars--Schneider integrable models from 5d \texorpdfstring{$\mathcal{N}=1$}{N=1} gauge theories}

\author{Gleb Arutyunov}
 \email{gleb.arutyunov@desy.de}
\author{Lukas Hardi}
 \email{lukas.hardi@desy.de}
\affiliation{
 II. Institut für Theoretische Physik, Universität Hamburg, \\ Luruper Chaussee 149, 22761 Hamburg, Germany
}

\date{\today}

\begin{abstract}
We show that the Coulomb branches of 5d $\mathcal{N}=1$ necklace quiver gauge theories on $\R^3 \times T^2$ are identified with the phase spaces of elliptic spin Ruijsenaars--Schneider models with dynamical inhomogeneities. This identification resolves the long-standing problem of determining a Poisson structure and a quantization of these models. We further show that the resulting quantum model is the integrable system governing supersymmetric indices of 4d $\mathcal{N}=1$ theories of class $\mathcal{S}_k$.
\end{abstract}

\maketitle
\setcounter{secnumdepth}{2}
\raggedbottom

\section{Introduction}\label{sec:intro}
The connection between supersymmetric gauge theories and integrable systems has a long history, beginning 
with the Seiberg--Witten solution of 4d $\mathcal{N}=2$ gauge theories on $\R^4$ \cite{seiberg:1994}
and the subsequent identification  of its integrable structure \cite{gorsky:1995b}. This framework was quickly extended to 3d $\mathcal{N}=4$ theories on $\R^3$
\cite{seiberg:1996,bullimore:2015} and 5d $\mathcal{N}=1$ theories on $\R^4 \times S^1$ \cite{nekrasov:1998b}. A mathematically rigorous definition of 3d $\mathcal{N}=4$ Coulomb branches was given by Braverman, Finkelberg, and Nakajima \cite{nakajima:2016}.
Their framework also applies to  
4d $\mathcal{N}=2$ theories compactified on 
$\R^3 \times S^1$, whose Coulomb branches are known as \emph{$K$-theoretic Coulomb branches} \cite{gaiotto:2010,schrader:2026}.

\begin{figure}[b]
    \centering
    \def\l{6}
    \def\radius{1.2cm}
    \begin{tikzpicture}[>=Stealth, every node/.style={font=\small}]
        \node[circle,draw,inner sep=1.5pt] (v1) at ({360*(0)/\l}:\radius) {$N$};
        \node[circle,draw,inner sep=1.5pt] (v2) at ({360*(1)/\l}:\radius) {$N$};
        \node[circle,draw,inner sep=1.5pt] (v3) at ({360*(2)/\l}:\radius) {$N$};
        \node[circle,draw,inner sep=1.5pt] (v4) at ({360*(3)/\l}:\radius) {$N$};
        \node (v5) at ({360*(4)/\l}:\radius) {$\cdots$};
        \node[circle,draw,inner sep=1.5pt] (v6) at ({360*(5)/\l}:\radius) {$N$};
        
        \foreach \i in {1,...,\l} {
            \pgfmathtruncatemacro{\next}{int(mod(\i,\l)+1)}
            \draw[->, shorten >=2pt, shorten <=2pt] (v\i) to[bend right=14] (v\next);
        }
        
        \node at ({(180+360*(0))/\l}:{1.2*\radius}) {\small $m^1$};
        \node at ({(180+360*(1))/\l}:{1.2*\radius}) {\small $m^2$};
        \node at ({(180+360*(2))/\l}:{1.2*\radius}) {\small $m^3$};
        \node at ({(180+360*(3))/\l}:{1.2*\radius}) {\small $m^4$};
        \node at ({(180+360*(4))/\l}:{1.2*\radius}) {\small $m^{\ell-1}$};
        \node at ({(180+360*(5))/\l}:{1.2*\radius}) {\small $m^0$};
    \end{tikzpicture}
    \caption{The necklace quiver with $\ell$ nodes of rank $N$ to which we associate a periodic quantum integrable spin chain. The bifundamental hypermultiplets corresponding to the arrows carry masses $m^\alpha$.}
    \label{fig:quiver}
\end{figure}
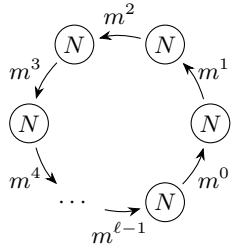

The 3d $\mathcal{N}=4$ and 4d $\mathcal{N}=2$ Coulomb branches on $\R^3$ and $\R^3 \times S^1$ are by now well understood. Treatments of their 5d $\mathcal{N}=1$ counterparts on $\R^3 \times T^2$, known as \emph{elliptic Coulomb branches} \cite{finkelberg:2022,webster:2026}, are more sparse. In this Letter, we examine the integrability of the Coulomb branch of 5d $\mathcal{N}=1$ necklace quiver theory on $\R^3 \times T^2$ with $\ell$ nodes of rank $N$, see Fig.~\ref{fig:quiver}, which we denote by $\mathcal{N}_{N,\ell}$ for short.
This theory has gauge group
\begin{equation}
    G = \prod_{\alpha \in \Z_\ell} \mathrm{U}(N)_\alpha
\end{equation}
and matter representation
\begin{equation}
    V = \bigoplus_{\alpha \in \Z_\ell} V_\alpha \otimes V_{\alpha+1}^\vee,
\end{equation}
where $\alpha \in \Z_\ell$ labels a node in the quiver and $V_\alpha$ is the vector representation of the gauge group factor $\mathrm{U}(N)_\alpha$, while $V_\alpha^\vee$ is its dual. The bifundamental hypermultiplet $V_\alpha \otimes V_{\alpha+1}^\vee$ is given the mass $m^\alpha$.

The Coulomb branches of the analogous 3d and 4d theories were previously identified with the phase spaces of the rational and trigonometric spin Ruijsenaars--Schneider models of $N$ particles and $\ell$ spin states with coupling constant $\eta = \sum_{\alpha \in \Z_\ell} m^\alpha$, respectively \cite{arutyunov:2026}. Our results imply the analogous identification between the Coulomb branch of $\mathcal{N}_{N,\ell}$ on $\R^3 \times T^2$ and the phase space of the elliptic spin Ruijsenaars--Schneider model. Moreover, the results of \cite{arutyunov:2026,arutyunov:2026b} are recovered by degenerating the underlying elliptic curve, thereby completing the 3d--4d--5d and rational--trigonometric--elliptic hierarchies.

The theory $\mathcal{N}_{N,\ell}$ also arises in the construction of 4d $\mathcal{N}=1$ theories of class $\mathcal{S}_k$ with $k = \ell$, obtained by compactifying the 6d $\mathcal{N}=(1,0)$ superconformal theory $\mathcal{T}_{N,\ell}$, defined by $N$ M5 branes probing an $A_{\ell-1}$ singularity, on a punctured Riemann surface \cite{gaiotto:2015}. Near a puncture, the  surface locally takes the form $\R_+ \times S^1$, and  reduction of $\mathcal{T}_{N,\ell}$ on $S^1$ is expected to yield $\mathcal{N}_{N,\ell}$ \cite{gaiotto:2015}. Gluing a trinion with two maximal and one minimal puncture corresponds to the insertion 
of a surface operator in the 4d $\mathcal{N}=1$ theory
and induces an action on its superconformal index by a family of commuting difference operators 
\cite{gaiotto:2015,maruyoshi:2016}.
In Sec.~\ref{sec:hams}, we identify these operators
with the Hamiltonians of the elliptic spin 
Ruijsenaars--Schneider model.

\subsection{5d \texorpdfstring{$\mathcal{N}=1$}{N=1} Coulomb branches}\label{sec:ellCB}

We use an abelianized description of the quantized algebra of functions on the 5d $\mathcal{N}=1$ Coulomb branch with quantization parameter $\hbar$, which we refer to as the \emph{abelianized Coulomb branch algebra}. This algebra quantizes the Poisson algebra introduced in  \cite{finkelberg:2022}, in direct analogy with the abelianized description of 3d $\mathcal{N}=4$ Coulomb branches \cite{bullimore:2015}. For the 5d $\mathcal{N}=1$ necklace quiver theory $\mathcal{N}_{N,\ell}$, it is generated by $\varphi_i^\alpha$ for $i=1,\dots,N$,  $\alpha \in \Z_\ell$, corresponding to the abelianized scalar fields in the vector multiplets, together with $u_i^{\alpha\pm}$, corresponding to the abelianized 't Hooft surface operators of minuscule magnetic charge wrapping $T^2$. These generators satisfy the relations \footnote{For the last two lines, there is an extra factor of $\theta(\eta)/\theta(\eta+\hbar)$ in the spinless case $\ell=1$.}
\begin{equation}
    \begin{aligned}
    [\varphi_i^\alpha,\varphi_j^\beta] ={}& 0, \\
    [\varphi_i^\alpha,u_j^{\beta\pm}] ={}& {\mp\hbar\delta_{ij}\delta^{\alpha\beta}}u_j^{\beta\pm}, \\
    u_i^{\alpha\pm} u_j^{\beta\pm} ={}& g_{\alpha \to \beta}^\pm(\varphi_{ij}^{\alpha\beta}) g_{\alpha = \beta}^\pm(\varphi_{ij}^{\alpha\beta}) g_{\alpha\leftarrow\beta}^\pm(\varphi_{ij}^{\alpha\beta}) u_j^{\beta\pm} u_i^{\alpha\pm}, \\
    u_i^{\alpha+} u_i^{\alpha-} ={}& \frac{\chi^{\alpha-1}(\varphi_i^\alpha+m^{\alpha-1}+\hbar) \chi^{\alpha+1}(\varphi_i^\alpha-m^\alpha)}{\prod_{k(\neq i)} \theta(\varphi_{ik}^{\alpha\alpha}+\hbar) \theta(\varphi_{ik}^{\alpha\alpha})}, \\
    u_i^{\alpha-} u_i^{\alpha+} ={}& \frac{\chi^{\alpha-1}(\varphi_i^\alpha+m^{\alpha-1}) \chi^{\alpha+1}(\varphi_i^\alpha-m^\alpha-\hbar)}{\prod_{k(\neq i)} \theta(\varphi_{ik}^{\alpha\alpha}) \theta(\varphi_{ik}^{\alpha\alpha}-\hbar)}
    \end{aligned}
\end{equation}
with $\varphi_{ij}^{\alpha\beta} \coloneq \varphi_i^\alpha - \varphi_j^\beta$,
\begin{equation}
    \chi^\alpha(z) \coloneq \prod_{i=1}^N \theta(z-\varphi_i^\alpha),
\end{equation}
and the structure factors
\begin{equation}
    \begin{aligned}
    g_{\alpha \to \beta}^\pm(\varphi_{ij}^{\alpha\beta}) &\coloneq g_{-1}^\pm(\varphi_{ij}^{\alpha\beta} - m^\alpha - \tfrac{\hbar}{2})^{\delta^{\alpha+1,\beta}}, \\
    g_{\alpha = \beta}^\pm(\varphi_{ij}^{\alpha\beta}) &\coloneq g_{2}^\pm(\varphi_{ij}^{\alpha\beta})^{\delta^{\alpha\beta}} ,\\
g_{\alpha\leftarrow\beta}^\pm(\varphi_{ij}^{\alpha\beta}) &\coloneq g_{-1}^\pm(\varphi_{ij}^{\alpha\beta} + m^\beta + \tfrac{\hbar}{2})^{\delta^{\alpha,\beta+1}},
    \end{aligned}
\end{equation}
defined in terms of
\begin{equation}
    g_\kappa^\pm(z) \coloneq \frac{\theta(z\mp \frac{\hbar \kappa}{2})}{\theta(z \pm \frac{\hbar \kappa}{2})}.
\end{equation}
Here $\theta(z)$ denotes the odd Jacobi theta function, with conventions specified in Sec.~\ref{sup:elliptic} of the Supplemental Material. Our aim is to identify the abelianized Coulomb branch algebra generated by $\varphi_i^\alpha$ and $u_i^{\alpha\pm}$ with the algebra of observables of the quantum elliptic spin Ruijsenaars--Schneider model. In the spinless case corresponding to $\ell=1$, this identification was previously established by supersymmetric localization \cite{yoshida:2021}.

\subsection{The spin Ruijsenaars--Schneider model}\label{sec:spinRS}

The elliptic spin Ruijsenaars--Schneider model is an integrable many-body system where each particle carries internal spin. Its degrees of freedom consist of particle positions $x_i$, $i=1,\dots,N$, together with spin covectors $a_i^\alpha$ and spin vectors $c_i^\alpha$, $\alpha \in \Z_\ell$. 
Krichever and Zabrodin originally formulated the model through its equations of motion and an $N\times N$ Lax pair \cite{krichever:1995}.
A symplectic form underlying these equations was subsequently constructed from the spectral decomposition of the Lax matrix \cite{krichever:2000}. An explicit Poisson algebra in the variables $x_i,a_i^\alpha,c_i^\alpha$, however, was accessible only in special cases \cite{arutyunov:1998,soloviev:2008,arutyunov:2026}. Our construction naturally yields a Poisson algebra, whose explicit form is given in Appendix~\ref{app:spinPoisson}.


The crucial step in obtaining a Poisson algebra is to deform the Krichever--Zabrodin equations of motion by introducing nonzero inhomogeneities $\Delta^\alpha$ for $\alpha \in \Z_\ell$:
\begin{equation}\label{eq:eoms}
    \begin{aligned}
        \dot x_i
        ={}& -\sum_{\rho \in \Z_\ell} f_{ii}^\rho, \\
        \dot a_i^\alpha
        ={}& \sum_{j=1}^N \sum_{\rho\in \Z_\ell} (V(\Delta^{0\rho},x_{ij}) f_{ij}^\rho a_i^\alpha - V(\Delta^{\alpha\rho},x_{ij}) f_{ij}^\rho a_j^\alpha), \\
        \dot c_i^\alpha
        ={}& \sum_{j=1}^N \sum_{\rho\in \Z_\ell} (V(\Delta^{\alpha\rho},x_{ji}) f_{ji}^\rho c_j^\alpha - V(\Delta^{0\rho},x_{ij}) f_{ij}^\rho c_i^\alpha).
    \end{aligned}
\end{equation}
Here $f_{ij}^\alpha \coloneq a_i^\alpha c_j^\alpha$, $\Delta^{\alpha\beta} \coloneq \Delta^\alpha - \Delta^\beta$, $x_{ij} \coloneq x_i-x_j$, and $a_i^0 = 1$, while the potential is given by 
\begin{equation}
    V(z,w) \coloneq \Phi(z,w) - \Phi(z,w-\eta),
\end{equation}
with $\eta$ the coupling constant and $\Phi(z,w)$ the Kronecker elliptic function, defined in Sec.~\ref{sup:elliptic} of the Supplemental Material. The Krichever--Zabrodin equations are recovered in the non-singular homogeneous limit $\Delta^\alpha \to 0$.  In this limit, the equations of motion acquire an enhanced spin-rotation symmetry. The corresponding symmetry algebra is the mirabolic subalgebra $\mathfrak{m}_\ell \subset \mathfrak{gl}_\ell$ preserving $a_i^0 = 1$, under which $a_i^\alpha$ and $c_i^\alpha$ transform as a covector and vector, respectively \cite{krichever:1995}. Nonzero inhomogeneities break the spin-rotation symmetry to its maximal torus, whose generators become identified with the inhomogeneities, see Appendix~\ref{app:spinPoisson}.

\section{Spin chain structure of the Coulomb branch}\label{sec:spinChain}

In this section, we identify the abelianized Coulomb branch algebra generated by $\varphi_i^\alpha$ and $u_i^{\alpha\pm}$ with the algebra of observables of a periodic quantum integrable spin chain whose sites correspond to the nodes of the necklace quiver. The key ingredients are  $N \times N$ $L$-operators associated with the arrows $\alpha \to \alpha+1$ in the necklace quiver, whose coefficients are defined by
\begin{equation}\label{eq:LOp}
    \begin{aligned}
        L_{ij}^{\alpha}(z) &\coloneq \Phi(z+\Delta^\alpha,\varphi_{ji}^{\alpha+1,\alpha}+m^\alpha) u_j^{\alpha+1,-},
    \end{aligned}
\end{equation}
where we have used $\Delta^\alpha \coloneq \sum_{i=1}^N (\varphi_i^\alpha-\varphi_i^{\alpha+1})$ as inhomogeneities. The difference operator representation of these $L$-operators (cf. Sec.~\ref{sup:GKLO} of the Supplemental Materials) is related to the $L$-operators of \cite{maruyoshi:2016}. In the spin-chain interpretation, $L^{\alpha}(z)$ transports excitations from site $\alpha+1$ to site $\alpha$. The monodromy matrix is therefore 
\begin{equation}
    L(z) \coloneq L^{0}(z) \cdots L^{\ell-1}(z)
\end{equation}
corresponding to clockwise transport around the spin chain.

Quantum integrability requires the transfer matrices $\operatorname{Tr} L(z)$ to commute for generic values of the spectral parameter.
To establish this property, we derive an $RLL$-type commutation relation for the $L$-operators.  


\subsection{\texorpdfstring{$L$}{L}-operator algebra}\label{sec:LOpAlg}

Using the ordering of $\varphi_i^\alpha$ and $u_i^{\alpha\pm}$ as in equation \eqref{eq:LOp}, we can derive the commutation relations of the $L$-operators $L^{\alpha}(z)$. These relations take an $RLL$-type form involving three $R$-matrices related to the $R$-matrices from \cite{arutyunov:1997}: $R^{\alpha\beta}(z,w)$ on the left, $\underline R^{\alpha\beta}(z-w)$ on the right, and $\bar R^{\alpha\beta}(z)$ sitting between the two $L$-operators. We now write the relations satisfied by these $R$-matrices in auxiliary space notation, where the subscripts 1, 2, and 3 denote auxiliary spaces of dimension $N$. The first $R$-matrix satisfies a version of the Yang--Baxter equation involving shifts of the spectral parameter:
\begin{align}\label{eq:shiftedYBE}
        R_{23}^{\beta\gamma}&(w+\hbar^{\alpha\beta},u+\hbar^{\alpha\gamma}) R_{13}^{\alpha\gamma}(z,u) R_{12}^{\alpha\beta}(z+\hbar^{\alpha\gamma},w+\hbar^{\beta\gamma})\nonumber \\
        &= R_{12}^{\alpha\beta}(z,w) R_{13}^{\alpha\gamma}(z+\hbar^{\alpha\beta},u+\hbar^{\beta\gamma}) R_{23}^{\beta\gamma}(w,u),
\end{align}
%
where $\hbar^{\alpha\beta} \coloneq \hbar \delta^{\alpha\beta}$. The second $R$-matrix $\underline R^{\alpha\beta}(z)$ coincides with Felder's elliptic dynamical $R$-matrix \cite{felder:1994b}, which satisfies the dynamical Yang--Baxter equation ~\\[-1pt]
\begin{equation}\label{eq:dynamicalYBE}
    \begin{aligned}
        (P_1^\alpha&)^{-1} \underline R_{23}^{\beta\gamma}(w-u) P_1^\alpha \underline R_{13}^{\alpha\gamma}(z-u) (P_3^\gamma)^{-1} \underline R_{12}^{\alpha\beta}(z-w) P_3^\gamma \\
        &= \underline R_{12}^{\alpha\beta}(z-w) (P_2^\beta)^{-1} \underline R_{13}^{\alpha\gamma}(z-u) P_2^\beta \underline R_{23}^{\beta\gamma}(w-u),
    \end{aligned}
\end{equation}
where $P^\alpha$ is the diagonal matrix whose $i$th entry is the operator which shifts the dynamical parameter $\varphi_i^\alpha$ by $\hbar$. Finally, $\bar R^{\alpha\beta}(z)$ is a spectral-parameter-dependent dynamical Drinfeld twist \cite{jimbo:1999} relating the two $R$-matrices 
through
\begin{equation}\label{eq:twist}
    R^{\alpha\beta}(z,w) = \bar R^{\alpha\beta}(z) \underline R^{\alpha\beta}(z-w) \bar R_{21}^{\beta\alpha}(w)^{-1}.
\end{equation}
The three relations are represented graphically in 
Fig.~\ref{fig:Rmatrelations}. Explicit expressions for the $R$-matrices are given in Sec.~\ref{sup:RMats} of the Supplemental Material.  Importantly, for $\alpha\neq \beta$, all three $R$-matrices reduce to the identity.

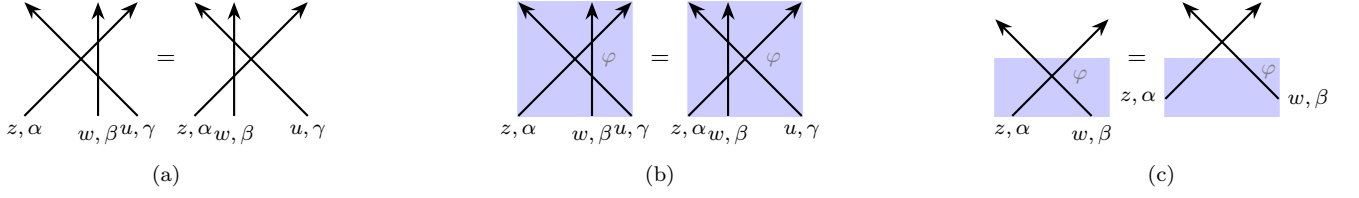
\begin{figure*}[t]
    \centering
    \subfloat[]{
    \begin{tikzpicture}[>=Stealth,thick,scale=0.75]
        \draw[->] (0.3,-1) -- (0.3,1);
        \draw[->] (-1,-1) -- (1,1);
        \draw[->] (1,-1) -- (-1,1);
        \node[below] at (-1,-1) {\footnotesize $z,\alpha$};
        \node[below] at (0.3,-1) {\footnotesize $w,\beta$};
        \node[below] at (1,-1) {\footnotesize $u,\gamma$};
        
        \node at (1.5,0) {$=$};
        
        \begin{scope}[xshift=3cm]
            \draw[->] (-0.3,-1) -- (-0.3,1);
            \draw[->] (-1,-1) -- (1,1);
            \draw[->] (1,-1) -- (-1,1);
            \node[below] at (-1,-1) {\footnotesize $z,\alpha$};
            \node[below] at (-0.3,-1) {\footnotesize $w,\beta$};
            \node[below] at (1,-1) {\footnotesize $u,\gamma$};
        \end{scope}
    \end{tikzpicture}
    }
    \hfill
    \subfloat[]{
    \begin{tikzpicture}[>=Stealth,thick,scale=0.75]
        \draw[blue!20!white,fill=blue!20!white] (-1,-1) rectangle (1,1);
        \draw[blue!20!white,fill=blue!20!white] (2,-1) rectangle (4,1);
    
        \draw[->] (0.3,-1) -- (0.3,1);
        \draw[->] (-1,-1) -- (1,1);
        \draw[->] (1,-1) -- (-1,1);
        \node[below] at (-1,-1) {\footnotesize $z,\alpha$};
        \node[below] at (0.3,-1) {\footnotesize $w,\beta$};
        \node[below] at (1,-1) {\footnotesize $u,\gamma$};
        \node[gray] at (0.6,0) {\footnotesize $\varphi$};
        
        \node at (1.5,0) {$=$};
        
        \begin{scope}[xshift=3cm]
            \draw[->] (-0.3,-1) -- (-0.3,1);
            \draw[->] (-1,-1) -- (1,1);
            \draw[->] (1,-1) -- (-1,1);
            \node[below] at (-1,-1) {\footnotesize $z,\alpha$};
            \node[below] at (-0.3,-1) {\footnotesize $w,\beta$};
            \node[below] at (1,-1) {\footnotesize $u,\gamma$};
            \node[gray] at (0.5,0) {\footnotesize $\varphi$};
        \end{scope}
    \end{tikzpicture}
    }
    \hfill
    \subfloat[]{
    \begin{tikzpicture}[>=Stealth,thick,scale=0.75]
        \draw[blue!20!white,fill=blue!20!white] (-1,-1) rectangle (1,0);
        \draw[blue!20!white,fill=blue!20!white] (2,-1) rectangle (4,0);
    
        \draw[->] (-0.7,-1) -- (1,0.7);
        \draw[->] (0.7,-1) -- (-1,0.7);
        \node[below] at (-0.7,-1) {\footnotesize $z,\alpha$};
        \node[below] at (0.7,-1) {\footnotesize $w,\beta$};
        \node[gray] at (0.5,-0.3) {\footnotesize $\varphi$};
        
        \node at (1.5,0) {$=$};

        \begin{scope}[xshift=3cm]
            \draw[->] (-1,-0.7) -- (0.7,1);
            \draw[->] (1,-0.7) -- (-0.7,1);
            \node[left] at (-1,-0.7) {\footnotesize $z,\alpha$};
            \node[right] at (1,-0.7) {\footnotesize $w,\beta$};
            \node[gray] at (0.82,-0.225) {\footnotesize $\varphi$};
        \end{scope}
    \end{tikzpicture}
    }
    \caption{Diagrammatic representation of the three relations satisfied by the $R$-matrices. (a) Shifted Yang--Baxter equation \eqref{eq:shiftedYBE} for the $R$-matrix $R^{\alpha\beta}(z,w)$, represented by a crossing on a white background. (b) Dynamical Yang--Baxter equation \eqref{eq:dynamicalYBE} for the dynamical $R$-matrix $\underline R^{\alpha\beta}(z-w)$, represented by a crossing on a blue background. Blue faces are labeled by the dynamical parameters $\varphi$, whose values are determined from the reference face, labeled $\varphi$, by the shift $\varphi^\alpha\to \varphi^{\alpha}+\hbar$ upon crossing a line labeled $\alpha$. (c) Twist relation \eqref{eq:twist} for the dynamical twist $\bar R^{\alpha\beta}(z)$, represented by two lines crossing from a blue to a white background. Composition is read from bottom to top.}
    \label{fig:Rmatrelations}
\end{figure*}

Combining these results yields the commutation relations \eqref{eq:LOpAlg} of the $L$-operators $L^{\alpha}(z)$.  In particular, the $L$-operators 
associated with nonadjacent arrows commute.
\begin{widetext}
\begin{align}
    R_{12}^{\alpha\beta} (z+\Delta^\alpha,w+\Delta^\beta) L_2^{\beta}(w) \bar R_{12}^{\alpha,\beta+1}(z+\Delta^\alpha)^{-1} L_1^{\alpha}(z) &= L_1^{\alpha}(z) \bar R_{21}^{\beta,\alpha+1}(w+\Delta^\beta)^{-1} L_2^{\beta}(w) \underline R_{12}^{\alpha+1,\beta+1}(z-w)\label{eq:LOpAlg} \\
    R_{12}^{00} (z+\Delta^0,w+\Delta^0) L_2(w) \bar R_{12}^{00}(z+\Delta^0)^{-1} L_1(z) &= L_1(z) \bar R_{21}^{00}(w+\Delta^0)^{-1} L_2(w) \underline R_{12}^{00}(z-w)\label{eq:LtotOpAlg} \\ \nonumber
\end{align}
\end{widetext}

\subsection{Commuting Hamiltonians}\label{sec:hams}

Repeated use of equations \eqref{eq:twist} and \eqref{eq:LOpAlg} yield equation \eqref{eq:LtotOpAlg}, which implies that the transfer matrix commutes for generic values of the spectral parameter by the argument from \cite{arutyunov:1997}:
\begin{equation}
    [\operatorname{Tr} L(z),\operatorname{Tr} L(w)] = 0.
\end{equation}
More generally, the transfer matrix belongs to a  family of Hamiltonians $S_n(z)$, defined as the coefficients of the quantum spectral curve
\begin{equation}\label{eq:qsc}
    {:\det(L(z)+\zeta \cdot \id) :} = \sum_{n=0}^N \zeta^{N-n} S_n(z),
\end{equation}
whose classical limit should be identified with the Seiberg--Witten curve of $\mathcal{N}_{N,\ell}$. We let $::$ denote the normal ordering in the difference operator representation, with conventions given in  Sec.~\ref{sup:GKLO} of the Supplemental Material. 
The first Hamiltonian is proportional to the transfer matrix
$S_1(z) \propto \operatorname{Tr} L(z)$ up to an unimportant prefactor. Explicit difference operator expressions for $S_n(z)$ are given in Appendix~\ref{app:diffOps}. For $\ell=1$, $S_n(z)$ reduces to the Hamiltonians of the elliptic spinless Ruijsenaars--Schneider model \cite{ruijsenaars:1987}, as expected from the transfer matrix construction of \cite{hasegawa:1995}. In general, for $n=1,\dots,N$,
$S_n(z)$ has $\ell$ simple poles whose residues furnish $N\ell$ difference operators. For $N,\ell \leq 3$, we verified explicitly that this family is commutative and functionally independent. Since the abelianized Coulomb branch algebra has $2N\ell$ functionally independent generators, these $N\ell$ commuting Hamiltonians establish Liouville integrability in the semiclassical limit.

For $N=2$, the Hamiltonian $S_1(z)$ 
can be compared directly with the commuting difference 
operators governing 
4d $\mathcal{N}=1$ class $\mathcal{S}_k$ theories with $k=\ell$ \cite{gaiotto:2015,maruyoshi:2016}.
In particular, $S_1(z)$ coincides with the $\mathrm{U}(2)$ analog of the $\mathrm{SU}(2)$-version appearing in \emph{loc.\ cit.}, see Appendix~\ref{app:diffOps}. In this sense, the operators $S_n(z)$ provide the higher-rank and higher-charge completion of the commuting difference operators governing 4d $\mathcal{N}=1$ theories of class $\mathcal{S}_k$ with $k=\ell$. In Sec.~\ref{sec:identification}, after an appropriate change of variables, we show that these commuting Hamiltonians coincide with those of the elliptic spin Ruijsenaars--Schneider model of Krichever and Zabrodin. Consequently, the elliptic spin Ruijsenaars--Schneider model with $\ell = k$ spin states emerges as the master integrable system governing supersymmetric indices of 4d $\mathcal{N}=1$ theories of class $\mathcal{S}_k$ of type $A$.

\section{Identification with spin Ruijsenaars--Schneider model}\label{sec:identification}

Passing to the semiclassical limit, the abelianized Coulomb branch Poisson algebra is identified with the Poisson algebra of the elliptic spin Ruijsenaars--Schneider model by the following identification of variables:
\begin{align}
    x_i &= \varphi_i^0 ,\label{eq:identA} \\
    a_i^\alpha &= e_i^t L^{0}(-\Delta^\alpha) \cdots L^{\alpha-1}(-\Delta^\alpha) e , \\
    c_i^\alpha &= u^{\alpha+1} L^{\alpha+1}(-\Delta^\alpha) \cdots L^{\ell-1}(-\Delta^\alpha) e_i, \label{eq:identB}
\end{align}
where $\alpha = 0,\dots,\ell-1$. Here $e_i$ denotes the $i$th standard basis vector and  $e \coloneq (1,\dots,1)^t$, while $u^\alpha \coloneq (u_1^{\alpha-},\dots,u_N^{\alpha-})$. Note that $a_i^0 = 1$. The inhomogeneities $\Delta^\alpha$, $\alpha \in \Z_\ell^\times$, generate the unbroken torus of the spin-rotation symmetry $\mathfrak{m}_\ell$:
\begin{equation}
    \{ \Delta^\alpha,a_i^\beta \} = \delta^{\alpha\beta} a_i^\beta, \qquad \{ \Delta^\alpha,c_i^\beta \} = -\delta^{\alpha\beta} c_i^\beta.
\end{equation}
Under this change of variables, the monodromy can be expressed in terms of $f_{ij}^\alpha = a_i^\alpha c_j^\alpha$:
\begin{equation}
    L_{ij}(z) = \sum_{\alpha \in \Z_\ell} \Phi(z+\Delta^\alpha,x_{ji}+\eta) f_{ij}^\alpha,
\end{equation}
with $\eta = \sum_{\alpha \in \Z_\ell} m^\alpha$. Consequently, the sum of residues of the transfer matrix yields
\begin{equation}\label{eq:ham}
    H \coloneq \sum_{i=1}^N \sum_{\alpha \in \Z_\ell} f_{ii}^\alpha,
\end{equation}
which belongs to the family of Poisson-commuting Hamiltonians. The Hamiltonian flow of $H$ 
reproduces the equations of motion \eqref{eq:eoms} of the elliptic spin Ruijsenaars--Schneider model.

\section{Conclusion}\label{sec:conclusion}

We have shown that the abelianized Coulomb branch algebra of the 5d $\mathcal{N}=1$ necklace quiver theory $\mathcal{N}_{N,\ell}$ on $\mathbb{R}^3 \times T^2$ is realized as the algebra of observables of a periodic quantum integrable spin chain of length $\ell$. The corresponding $N \times N$ $L$-operators have coefficients in the abelianized Coulomb branch algebra \eqref{eq:LOp} and satisfy an $RLL$-type commutation relation \eqref{eq:LOpAlg}. The latter involves a spectral-parameter-dependent $R$-matrix $R^{\alpha\beta}(z,w)$, Felder's elliptic dynamical $R$-matrix $\underline{R}^{\alpha\beta}(z)$, and a dynamical Drinfeld twist $\bar{R}^{\alpha\beta}(z)$. The coefficients $S_n(z)$ of the quantum spectral curve \eqref{eq:qsc} form a commuting family, which is identified through the change of variables \eqref{eq:identA}--\eqref{eq:identB} with the Hamiltonians of the elliptic spin Ruijsenaars--Schneider model of Krichever and Zabrodin in the sense that the Hamiltonian flow generated by $H$ from \eqref{eq:ham} reproduces the equations of motion \eqref{eq:eoms}. The Coulomb branch construction therefore provides the missing components for the model: an explicit Poisson algebra in the natural variables $x_i,a^\alpha_i,c^\alpha_i$, given in Appendix~\ref{app:spinPoisson}, together with its quantization.

Taking $N=2$ and imposing $\mathrm{SU}(2)$ constraints, the transfer matrix $S_1(z)$ reduces to the commuting difference operators \eqref{eq:diffOpsSU2} acting on the supersymmetric index of $4$d $\mathcal{N}=1$ class $\mathcal{S}_k$ with $k=\ell$ \cite{maruyoshi:2016}. In this sense, the elliptic spin Ruijsenaars--Schneider model provides   the integrable structure underlying class $\mathcal{S}_k$, analogous to the role of the elliptic (spinless) Ruijsenaars--Schneider model for class $\mathcal{S}$ \cite{gaiotto:2013}. The operators $S_n(z)$ constructed here furnish the higher-rank and higher-charge completion of the $\mathrm{SU}(2)$ operators of \cite{gaiotto:2015,maruyoshi:2016}.

Future directions include clarifying the relation between the commuting difference operators $S_n(z)$,  the spin-DELL Hamiltonians of \cite{koroteev:2020b}, and the matrix-valued difference operators of \cite{klabbers:2024b}. It would also be important to develop Hilbert space representations. In the trigonometric degeneration, the framework of Schur quantization \cite{gaiotto:2024b} may provide a natural route to such a construction.

\begin{acknowledgments}
\paragraph{Acknowledgments.} We thank Samuel DeHority, Craig Lawrie, Elli Pomoni, Shlomo Razamat, Volker Schomerus, Mykola Semenyakin, and Mayuko Yamashita for helpful discussions. G.A. acknowledges support by the DFG under Germany's Excellence Strategy -- EXC 2121 ``Quantum Universe'' -- 390833306. G.A. and L.H. acknowledge support by the DFG -- SFB 1624 -- ``Higher structures, moduli spaces and integrability'' -- 506632645. This research was supported in part by Perimeter Institute for Theoretical Physics. Research at Perimeter Institute is supported by the Government of Canada through the Department of Innovation, Science and Economic Development and by the Province of Ontario through the Ministry of Research, Innovation and Science.
\end{acknowledgments}

\appendix

\section{Commuting difference operators}\label{app:diffOps}

Here we give the explicit difference operator realization of the coefficients $S_n(z)$ of the quantum spectral curve
\begin{equation}
    {:\det(L(z)+\zeta \cdot \id) :} = \sum_{n=0}^N \zeta^{N-n} S_n(z),
\end{equation}
where normal ordering is understood in the difference operator representation (cf. Sec.~\ref{sup:GKLO} of the Supplemental Material). Introducing the compensating sum
\begin{equation}
    \Sigma_I^\alpha \coloneq n m^{\alpha-1} - \sum_{i \notin I_\alpha} \varphi_i^\alpha + \sum_{i \notin I_{\alpha-1}} \varphi_i^{\alpha-1},
\end{equation}
the coefficients are explicitly given by \eqref{eq:diffOps}. For $N=2$, setting $\varphi^\alpha \coloneq \tfrac{1}{2}(\varphi_1^\alpha-\varphi_2^\alpha)$ and imposing the $\mathrm{SU}(2)$ constraints $\varphi_1^\alpha + \varphi_2^\alpha = 0$, equation \eqref{eq:diffOps} reduces to 
\eqref{eq:diffOpsSU2}. This coincides with the commuting difference operators governing the 4d $\mathcal{N}=1$ theories of class $\mathcal{S}_k$ with $k=\ell$ \cite{gaiotto:2015,maruyoshi:2016} at specific values of the spectral parameter $z$.
\begin{widetext}
\begin{equation}\label{eq:diffOps}
    S_n(z) = \sum_{\substack{I_0,\dots,I_{\ell-1} \subseteq \{ 1,\dots,N \} \\ |I_\alpha|=n}} \prod_{\alpha \in \Z_\ell} \frac{\theta(z+\Sigma_I^\alpha)}{\theta(z+\Delta^\alpha)} \frac{\prod_{i \in I_\alpha,k \notin I_{\alpha-1}} \theta(\varphi_i^\alpha-\varphi_k^{\alpha-1}+m^{\alpha-1})}{\prod_{i \in I_{\alpha},k \notin I_\alpha} \theta(\varphi_i^\alpha-\varphi_k^\alpha)} \prod_{\alpha \in \Z_\ell} \prod_{i \in I_\alpha} e^{-\hbar\partial_{\varphi_i^\alpha}}
\end{equation}
\begin{equation}\label{eq:diffOpsSU2}
    S_1(z) \prod_{\alpha\in\Z_\ell} e^{\hbar (\partial_{\varphi_1^\alpha} + \partial_{\varphi_2^\alpha})} = \sum_{(s^\alpha) \in \{ \pm 1 \}^\ell} \prod_{\alpha \in \Z_\ell} \frac{\theta(z + s^\alpha \varphi^\alpha - s^{\alpha-1} \varphi^{\alpha-1} + m^{\alpha-1})}{\theta(z)} \frac{\theta(s^\alpha \varphi^\alpha + s^{\alpha-1}\varphi^{\alpha-1} + m^{\alpha-1})}{\theta(2 s^\alpha \varphi^\alpha)} \prod_{\alpha \in \Z_\ell} e^{-s^\alpha \hbar \partial_{\varphi^\alpha}}
\end{equation}
\end{widetext}

\section{Poisson bracket of spin variables}\label{app:spinPoisson}

Below, we give the Poisson algebra of the elliptic spin Ruijsenaars--Schneider model in terms of the variables $x_i,a_i^\alpha,c_i^\alpha,\Delta^\alpha$, with $i=1,\dots,N$ and $\alpha \in \Z_\ell$, introduced in Sec.~\ref{sec:spinRS}. It follows from  the difference operator representation given in  Sec.~\ref{sup:GKLO} of the Supplemental Material.
\begin{align}
    \{ a_i^\alpha,a_j^\beta \}
    ={}& E_1(x_{ij}) a_i^\alpha a_j^\beta \nonumber \\
    &- \Phi(\Delta^{\beta\alpha},x_{ji}) a_j^\alpha a_i^\beta \nonumber \\
    &+ \Phi(\Delta^{0\alpha},x_{ji}) a_j^\alpha a_j^\beta \nonumber \\
    &- \Phi(\Delta^{0\beta},x_{ij}) a_i^\alpha a_i^\beta, \displaybreak \\
    \{ a_i^\alpha,c_j^\beta \}
    ={}& {-E_1(x_{ij})} a_i^\alpha c_j^\beta \nonumber \\ 
    &- \Phi(\Delta^{0\alpha},x_{ji}) a_j^\alpha c_j^\beta \nonumber \\
    &+ \delta^{\alpha\beta} L_{ij}(-\Delta^\alpha) \nonumber \\
    &- \delta^{0\beta} L_{ij}(-\Delta^0) a_i^\alpha, \\
    \{ c_i^\alpha,c_j^\beta \}
    ={}& E_1(x_{ij}) c_i^\alpha c_j^\beta \nonumber \\ 
    &+ \Phi(\Delta^{\beta\alpha},x_{ij}) c_j^\alpha c_i^\beta \nonumber \\
    &- \delta^{\alpha0} L_{ji}(-\Delta^0) c_j^\beta \nonumber \\
    &+ \delta^{0\beta} L_{ij}(-\Delta^0) c_i^\alpha, \\
    \{ x_i,a_j^\alpha \}
    ={}& 0, \\ 
    \{ x_i,c_j^\alpha \}
    ={}& \delta_{ij} c_j^\alpha, \\
    \{ \Delta^\alpha,a_i^\beta \}
    ={}& \delta^{\alpha\beta} a_i^\beta, \qquad (\alpha \in \Z_\ell^\times) \\ 
    \{ \Delta^\alpha,c_i^\beta \}
    ={}& {-\delta^{\alpha\beta}} c_i^\beta, \quad \ (\alpha \in \Z_\ell^\times) \\
    \{ \Delta^\alpha,x_i \}
    ={}& 0, \\
    \{ x_i,x_j \}
    ={}& 0.
\end{align}
One may check that this Poisson bracket is antisymmetric and satisfies the Jacobi identity independently of the representation in terms of the Coulomb branch variables upon imposing the constraints $a_i^0 = 1$ and $\Delta^0 = -\sum_{\alpha \in \Z_\ell^\times} \Delta^\alpha$, leaving $2 N \ell + (\ell-1)$ independent generators. The Jacobi identity relies crucially on the fact that $\Delta^\alpha$ does not Poisson-commute with $a_i^\alpha$ and $c_i^\alpha$. A calculation for small cases  ($N,\ell \leq 3$) shows that the Poisson bivector has rank $2N \ell$ and thus the Poisson bracket possesses $\ell-1$ Casimirs.

The variables $\Delta^\alpha$ for $\alpha \in \Z_\ell^\times$ form the moment map for the action of the torus of the spin-rotation symmetry algebra $\mathfrak{m}_\ell$. We may therefore perform the Hamiltonian reduction with respect to this moment map by restricting to the torus invariants $f_{ij}^\alpha \coloneq a_i^\alpha c_j^\alpha$ and fixing the level set of $\Delta^\alpha$. The resulting Poisson bracket of the torus invariants $f_{ij}^\alpha$ is given by
\begin{equation}\label{eq:invbracket}
    \begin{aligned}
        \{ f_{ij}^\alpha, f_{kl}^\beta \}
        ={}& (E_1(x_{ik})+E_1(x_{jl})-E_1(x_{il})-E_1(x_{jk})) f_{ij}^\alpha f_{kl}^\beta \\
        &+ (\Phi(\Delta^{\beta\alpha},x_{jl}) f_{il}^\alpha f_{kj}^\beta - \Phi(\Delta^{\beta\alpha},x_{ki}) f_{kj}^\alpha f_{il}^\beta) \\
        &+ (\Phi(\Delta^{0\alpha},x_{ki}) f_{kj}^\alpha - \Phi(\Delta^{0\alpha},x_{li}) f_{lj}^\alpha) f_{kl}^\beta \\
        &+ f_{ij}^\alpha (\Phi(\Delta^{0\beta},x_{jk}) f_{jl}^\beta - \Phi(\Delta^{0\beta},x_{ik}) f_{il}^\beta) \\
        &+ \delta^{\alpha\beta} (L_{il}(-\Delta^\alpha) f_{kj}^\alpha - L_{kj}(-\Delta^\alpha) f_{il}^\alpha) \\
        &+ \delta^{\alpha0} (L_{kj}(-\Delta^0)-L_{lj}(-\Delta^0)) f_{kl}^\beta \\
        &+ \delta^{0\beta} f_{ij}^\alpha (L_{jl}(-\Delta^0)-L_{il}(-\Delta^0)).
    \end{aligned}
\end{equation}
This bracket satisfies the Jacobi identity upon imposing the rank-one condition $f_{ij}^\alpha = a_i^\alpha c_j^\alpha$. Unlike the equations of motion \eqref{eq:eoms}, however, the bracket \eqref{eq:invbracket} is singular in the homogeneous limit. This reflects the enhanced spin-rotation symmetry in this limit, under which $f_{ij}^\alpha$ is not invariant. Considering instead the full invariants $f_{ij} \coloneq \sum_{\alpha \in \Z_\ell} f_{ij}^\alpha$, we find that their bracket does not close unless the homogeneous limit is taken, which is non-singular, giving
\begin{equation}\label{eq:ffbracket}
    \begin{aligned}
        \{ f_{ij},f_{kl} \}
        ={}& (E_1(x_{ik})+E_1(x_{jl})-E_1(x_{il})-E_1(x_{jk})) f_{ij} f_{kl} \\
        &+ (E_1(x_{jl}) f_{il} f_{kj} - E_1(x_{ki}) f_{kj} f_{il}) \\
        &+ (E_1(x_{ki}) f_{kj} - E_1(x_{li}) f_{lj}) f_{kl} \\
        &+ f_{ij} (E_1(x_{jk}) f_{jl} - E_1(x_{ik}) f_{il}) \\
        &+ (E_1(\eta+x_{li}) f_{il} f_{kj} - E_1(\eta+x_{jk}) f_{kj} f_{il}) \\
        &+ (E_1(\eta+x_{jk}) f_{kj}-E_1(\eta+x_{jl}) f_{lj}) f_{kl} \\
        &+ f_{ij} (E_1(\eta+x_{lj}) f_{jl}-E_1(\eta+x_{li}) f_{il}).
    \end{aligned}
\end{equation}
One may check that this bracket does not satisfy the Jacobi identity, although its trigonometric and rational degeneration do \cite{arutyunov:1998}. The reason is that, near  the homogeneous limit, the Jacobiator 
admits 
an $\epsilon$-expansion of the schematic form
\begin{equation}
    \begin{aligned}
    J(f_{ij},f_{kl},f_{mn}) ={}& \{ \eqref{eq:ffbracket}, f_{mn} \} + \epsilon \{ c_{ijkl}, f_{mn} \} + O(\epsilon^2) \\
    &+ \text{cycl.},
    \end{aligned}
\end{equation}
where $c_{ijkl}$ can only be expressed in terms of the individual $f_{ij}^\alpha$. Their Poisson brackets generate singular terms that combine with the explicit factor of $\epsilon$ in front of $c_{ijkl}$ and critically contribute to the vanishing of the Jacobiator in the homogeneous limit. These terms are not visible from the homogeneous bracket \eqref{eq:ffbracket}, which therefore fails to satisfy the Jacobi identity. Thus, at the elliptic level, the Poisson structure crucially relies on the presence of dynamical inhomogeneities.

The Poisson bracket of the monodromy $L(z)$ coming from \eqref{eq:LtotOpAlg} implies the existence of a Lax pair satisfying the Lax equation
\begin{equation}\label{eq:lax}
    \{ H,L(z) \} = [M(z),L(z)].
\end{equation}
Concretely,
\begin{equation}
    \begin{aligned}
        M(z) ={}& \sum_{i=1}^N \sum_{\alpha \in \Z_\ell} E_1(z+\Delta^\alpha) f_{ii}^\alpha E_{ii} \\
        &+ \sum_{i,j=1}^N \sum_{\alpha \in \Z_\ell} V(\Delta^{0\alpha},x_{ij}) f_{ij}^\alpha E_{ii} \\
        &+ \sum_{i \neq j}^N \sum_{\alpha \in \Z_\ell} \Phi(z+\Delta^\alpha,x_{ji}) f_{ij}^\alpha E_{ij}.
    \end{aligned}
\end{equation}
Like the equations of motion \eqref{eq:eoms}, the inhomogeneous Lax equation \eqref{eq:lax} admits a non-singular homogeneous limit. In this limit, $L(z)$ and $M(z)$ reduce precisely to the Lax pair of Krichever and Zabrodin \cite{krichever:1995}, from which the symplectic form of the elliptic spin Ruijsenaars--Schneider model was constructed \cite{krichever:2000}. Unlike the homogeneous Lax pair, the inhomogeneous one resolves distinct spin states and generates the equations of motion for the individual spins.

\pagebreak
~
\pagebreak


\renewcommand{\thesection}{S}
\renewcommand{\thesubsection}{\arabic{subsection}}
\renewcommand{\theequation}{S\arabic{equation}}
\renewcommand{\thefigure}{S\arabic{figure}}
\renewcommand{\thetable}{S\arabic{table}}
\setcounter{equation}{0}
\setcounter{figure}{0}
\setcounter{table}{0}

\onecolumngrid
\section*{Supplemental Material}
\twocolumngrid

\subsection{Index of notations}\label{sup:notations}

\begin{center}
    \setlength{\tabcolsep}{4pt} 
    \renewcommand{\arraystretch}{1.25} 
    \begin{tabular}{|l|l|}
        \hline
        $\ell$ & number of nodes in the necklace quiver \\
        $\Z_\ell$ & integers modulo $\ell$ \\
        $\Z_\ell^\times$ & $\Z_\ell \setminus \{ 0 \}$ \\
        $N$ & rank of gauge group factors \\
        $\mathcal{N}_{N,\ell}$ & 5d $\mathcal{N}=1$ necklace quiver theory \\
        $\mathrm{U}(N)_\alpha$ & $\alpha$th gauge group factor of $\mathcal{N}_{N,\ell}$ \\
        $V_\alpha$ & vector representation of $\mathrm{U}(N)_\alpha$ \\
        $V_\alpha^\vee$ & dual vector representation of $\mathrm{U}(N)_\alpha$ \\
        $m^\alpha$ & mass of bifundamental $V_\alpha \otimes V_{\alpha+1}^\vee$ \\
        $\mathcal{T}_{N,\ell}$ & 6d $\mathcal{N}=(1,0)$ SCFT of $N$ M5 branes \\
        & placed at an $A_{\ell-1}$ singularity \\
        $\varphi_i^\alpha$ & abelianized vector multiplet scalar \\
        $\varphi_{ij}^{\alpha\beta}$ & $\varphi_i^\alpha - \varphi_j^\beta$ \\
        $u_i^{\alpha\pm}$ & abelianized 't Hooft surface operators \\
        & of minuscule magnetic charge \\
        $\theta(z)$ & odd theta function \eqref{eq:oddTheta} \\
        $\chi^\alpha(z)$ & $\prod_{i=1}^N \theta(z-\varphi_i^\alpha)$ \\
        $x_i$ & particle positions \\
        $a_i^\alpha,c_i^\alpha$ & $i$th spin covector and vector \\
        $\Delta^\alpha$ & inhomogeneity, $\sum_{i=1}^N (\varphi_i^\alpha-\varphi_i^{\alpha+1})$ \\
        $f_{ij}^\alpha$ & $a_i^\alpha c_j^\alpha$ \\
        $f_{ij}$ & $\sum_{\alpha \in \Z_\ell} f_{ij}^\alpha$ \\
        $\Delta^{\alpha\beta}$ & $\Delta^\alpha - \Delta^\beta$ \\
        $x_{ij}$ & $x_i - x_j$ \\
        $\Phi(z,w)$ & Kronecker elliptic function \eqref{eq:kronecker} \\
        $\eta$ & spin RS coupling constant \\
        $V(z,w)$ & potential $\Phi(z,w) - \Phi(z,w-\eta)$ \\
        $L^{\alpha}(z)$ & $L$-operators associated to $\alpha \to \alpha+1$ \\
        $L(z)$ & monodromy $L^{0}(z) \cdots L^{\ell-1}(z)$ \\
        $R^{\alpha\beta}(z,w)$ & $R$-matrix satisfying \eqref{eq:shiftedYBE} \\
        $\underline R^{\alpha\beta}(z)$ & Felder's $R$-matrix satisfying \eqref{eq:dynamicalYBE} \\
        $\bar R^{\alpha\beta}(z)$ & Drinfeld twist satisfying \eqref{eq:twist} \\
        $S_n(z)$ & commuting difference operators \\ & from quantum spectral curve \eqref{eq:qsc} \\
        $H$ & spin RS Hamiltonian \eqref{eq:ham} \\
        \hline
    \end{tabular}
\end{center}

\subsection{Elliptic conventions}\label{sup:elliptic}

Throughout the text, we consider the elliptic curve $E_\tau = \C/(\Z + \tau \Z)$ and let $\theta \in \mathcal{O}(1)$ be
\begin{equation}\label{eq:oddTheta}
    \theta(z) \coloneq \frac{\theta_1(\pi z,e^{\I \pi \tau})}{\pi \theta_1'(0,e^{\I \pi \tau})},
\end{equation}
which satisfies the quasi-periodicity
\begin{equation}
    \theta(z+1) = -\theta(z), \quad \theta(z+\tau) = -e^{-\I\pi \tau - 2\pi \I z} \theta(z),
\end{equation}
and is normalized such that $\theta'(0) = 1$. We also use its logarithmic derivative
\begin{equation}
    E_1(z) \coloneq \frac{\theta'(z)}{\theta(z)},
\end{equation}
which has a simple pole at $z=0$ with residue one. We take the regular part and set $E_1(0) \coloneq 0$. Furthermore, we make use of the Kronecker elliptic function
\begin{equation}\label{eq:kronecker}
    \Phi(z,u) \coloneq \frac{\theta(z+u)}{\theta(z) \theta(u)},
\end{equation}
which is symmetric $\Phi(z,u) = \Phi(u,z)$, satisfies the quasi-periodicity
\begin{equation}
    \Phi(z+1,u) = \Phi(z,u), \quad \Phi(z+\tau,u) = e^{-2\pi\I u} \Phi(z,u),
\end{equation}
and has a simple pole at $z=0$ with residue one. We again take the regular part and set $\Phi(z,0) = \Phi(0,z) \coloneq E_1(z)$.

\subsection{Explicit form of \texorpdfstring{$R$}{R}-matrices}\label{sup:RMats}

Let us list the explicit form of the $R$-matrices from equations \eqref{eq:LOpAlg} and \eqref{eq:LtotOpAlg} expressed in terms of $N \times N$ matrix units $e_{ij}$:
\begin{align}
        R^{\alpha\alpha}(z,w) \coloneq{}& \sum_{i,j=1}^N \frac{\Phi(\hbar,\varphi_{ij}^{\alpha\alpha})}{\Phi(w-z,\hbar)} e_{ii} \otimes e_{jj} \\
        &+ \sum_{i,j=1}^N \frac{\Phi(w-z,\varphi_{ij}^{\alpha\alpha})}{\Phi(w-z,\hbar)} e_{ij} \otimes e_{ji} \nonumber \\
        &- \sum_{i,j=1}^N \frac{\Phi(-z,\varphi_{ij}^{\alpha\alpha})}{\Phi(w-z,\hbar)} e_{ij} \otimes e_{jj} \nonumber \\
        &+ \sum_{i,j=1}^N \frac{\Phi(-w-\hbar,\varphi_{ij}^{\alpha\alpha})}{\Phi(w-z,\hbar)} e_{jj} \otimes e_{ij}, \nonumber \displaybreak \\
        \bar R^{\alpha\alpha}(z) \coloneq{}& \sum_{i,j=1}^N \theta(\hbar) \Phi(\hbar,\varphi_{ij}^{\alpha\alpha}-\hbar) e_{ii} \otimes e_{jj} \\
        &- \sum_{i,j=1}^N \theta(\hbar) \Phi(-z,\varphi_{ij}^{\alpha\alpha}-\hbar) e_{ij} \otimes e_{jj}, \nonumber \\
        \underline R^{\alpha\alpha}(z) \coloneq{}& \sum_{i \neq j}^N \frac{\Phi(\hbar,\varphi_{ji}^{\alpha\alpha})}{\Phi(-z,\hbar)} e_{ii} \otimes e_{jj} +  \sum_i e_{ii} \otimes e_{ii} \nonumber \\
        &- \sum_{i \neq j}^N \frac{\Phi(z,\varphi_{ji}^{\alpha\alpha})}{\Phi(-z,\hbar)} e_{ij} \otimes e_{ji},
\end{align}
We set $R^{\alpha\beta}(z,w)$, $\bar R^{\alpha\beta}(z)$, and $\underline R^{\alpha\beta}(z)$ equal to the identity whenever $\alpha \neq \beta$.

\subsection{\label{sup:GKLO}Elliptic GKLO representation}

The abelianized Coulomb branch algebra, whose relations are written down in Sec.~\ref{sec:ellCB}, can be represented by difference operators. This representation is an elliptic generalization of the GKLO representation \cite{gerasimov:2005}. Concretely, the abelianized 't Hooft surface operators $u_i^{\alpha\pm}$ are represented as follows:
\begin{equation}
    \begin{aligned}
        u_i^{\alpha+} &\mapsto \frac{\chi^{\alpha+1}(\varphi_i^\alpha-m^\alpha)}{\prod_{k(\neq i)} \theta(\varphi_{ik}^{\alpha\alpha})} e^{\hbar\partial_{\varphi_i^\alpha}} \\
        u_i^{\alpha-} &\mapsto \frac{\chi^{\alpha-1}(\varphi_i^\alpha+m^{\alpha-1})}{\prod_{k(\neq i)} \theta(\varphi_{ik}^{\alpha\alpha})} e^{-\hbar\partial_{\varphi_i^\alpha}}.
    \end{aligned}
\end{equation}
One can check that this representation indeed obeys the relations from Sec.~\ref{sec:ellCB}. This difference operator representation is the main algebraic tool for recovering the formulas discussed in this Letter. Whenever we write $::$, we mean the normal ordering in the difference operator representation which orders all shift operators $e^{\pm\hbar\partial_{\varphi_i^\alpha}}$ to the right.
\end{document}